\documentclass[letterpaper, 11pt]{article}
\usepackage{latexsym}
\usepackage{amssymb}
\usepackage{times}
\usepackage{amsmath,amsfonts,amsthm}
\usepackage{graphicx}
\usepackage{algpseudocode}
\usepackage{algorithm}

\newcommand{\ble}{\begin{lemma}}
\newcommand{\ele}{\end{lemma}}

\newtheorem{lemma}{Lemma}[section]

\newtheorem{theorem}[lemma]{Theorem}

\newtheorem{definition}[lemma]{Definition}

\newtheorem{fact}[lemma]{Fact}

\newtheorem{algorithm1}[lemma]{Algorithm}

\newcommand{\beao}{\begin{eqnarray*}}
\newcommand{\eeao}{\end{eqnarray*}\noindent}
\newcommand{\beam}{\begin{eqnarray}}
\newcommand{\eeam}{\end{eqnarray}\noindent}

\newcommand{\one}{{\bf 1}}

\begin{document}

\title{Approximating Large Frequency Moments with Pick-and-Drop Sampling}
\author{Vladimir Braverman\thanks{Johns Hopkins University, vova@cs.jhu.edu},\ \  Rafail Ostrovsky\thanks{University of California Los Angeles,
Department of Computer Science and Department of Mathematics,
Email: rafail@cs.ucla.edu.
Research supported in part
by NSF grants CCF-0916574; IIS-1065276; CCF-1016540; CNS-1118126;
CNS-1136174; US-Israel BSF grant
2008411, OKAWA Foundation Research Award, IBM Faculty Research Award, Xerox
Faculty Research Award, B. John
Garrick Foundation Award, Teradata Research Award, and Lockheed-Martin
Corporation Research Award. This material is
also based upon work supported by the Defense Advanced Research Projects
Agency through the U.S. Office of Naval Research
under Contract N00014-11-1-0392. The views expressed are those of the author
and do not reflect the official policy or position
of the Department of Defense or the U.S. Government.}\\}

\maketitle

\begin{abstract}
Given data stream $D = \{p_1,p_2,\dots,p_m\}$ of
size $m$ of numbers from $\{1,\dots, n\}$, the frequency of $i$ is
defined as $f_i = |\{j: p_j = i\}|$. The $k$-th \emph{frequency
moment} of $D$ is defined as $F_k = \sum_{i=1}^n f_i^k$.
We consider the problem of approximating frequency moments in insertion-only streams for $k\ge 3$. For any constant $c$ we show an $O(n^{1-2/k}\log(n)\log^{(c)}(n))$ upper bound on the space complexity of the problem. Here $\log^{(c)}(n)$ is the iterative $\log$ function.
To simplify the presentation, we make the following assumptions: $n$ and $m$
are polynomially far; approximation error $\epsilon$ and parameter $k$ are
constants.
We observe a natural bijection between streams and special matrices.
Our main technical contribution is a non-uniform sampling method on matrices. We call our method a \emph{pick-and-drop
sampling}; it samples a heavy element (i.e., element $i$ with frequency $\Omega(F_k)$) with probability $\Omega(1/n^{1-2/k})$ and gives approximation $\tilde{f_i} \ge (1-\epsilon)f_i$. In addition, the estimations never exceed the real values, that is $ \tilde{f_j} \le f_j$ for all $j$. As a result, we reduce the space complexity of finding a heavy element to $O(n^{1-2/k}\log(n))$ bits.
We apply our method of recursive sketches and resolve the problem with $O(n^{1-2/k}\log(n)\log^{(c)}(n))$ bits. \end{abstract}

\section{Introduction}

Given a sequence $D = \{p_1,p_2,\dots,p_m\}$ of
size $m$ of numbers from $\{1,\dots, n\}$, a frequency of $i$ is
defined as
\begin{equation}\label{kljdflkjsdfljdsfljdsfklj}
f_i = |\{j: p_j = i\}|.
\end{equation}
 The $k$-th \emph{frequency
moment} of $D$ is defined as
\begin{equation}\label{kljdfdsdfddflkjsdfljdsfljdsfklj}
F_k = \sum_{i=1}^n f_i^k.
\end{equation}

The problem of approximating frequency moments in one pass over $D$
and using \emph{sublinear} space has been introduced in the
award-winning paper of Alon, Matias and Szegedy \cite{ams}. In
particular, they observed a striking difference between
``small'' and ``large'' values of $k$: it is possible
to approximate $F_k, k\le 2$ in polylogarithmic space, but
polynomial space is required when $k>2$. Since $1996$,
approximating $F_k$ has become one of the most
inspiring problems in the theory of data streams. The incomplete
list of papers on frequency moments include
 \cite{stable,5215,
frequency_lower_bound1, frequency_lower_bound2, 711822, frequency,
frequency_impr2, 776778, 796530, frequency_impr1, 1459774, 1496816, recursive,
nelson, 1807094, 1807101, 982817, 1250891, 1374470, 1265565, DBLP:conf/focs/AndoniKO11, DBLP:journals/corr/abs-1104-4552, DBLP:journals/corr/abs-1201-0253, Woodruff:2012:TBD:2213977.2214063, Jayram:2011:OBJ:2133036.2133037} and
references therein.
We omit the detailed history of the problem and refer a reader to \cite{strbook, DBLP:reference/db/Woodruff09} for overviews.

In this paper we consider the case when $k\ge 3$.
In their breakthrough paper Indyk and Woodruff \cite{frequency} gave the first solution that is optimal up to a polylogarithmic factor. Numerous improvements were proposed in the later years (see the references above) and the latest bounds are due to Andoni, Krauthgamer and Onak \cite{DBLP:conf/focs/AndoniKO11} and Ganguly \cite{DBLP:journals/corr/abs-1104-4552}.
The latest bound by Ganguly \cite{DBLP:journals/corr/abs-1104-4552} is
$$ O(k^2\epsilon^{-2}n^{1-2/k}E(p,n) \log (n) \log (nmM)/\min(\log (n),\epsilon^{4/k-2}))$$ where, $E(k,n) = (1-2/k)^{-1}(1-n^{-4(1-2/k})$. This bound is roughly $O(n^{1-2/k}\log^2(n))$ for constant $\epsilon, k$.
The best known lower bound for insertion-only streams is $\Omega(n^{1-2/k})$, due to Chakrabarti, Khot and Sun \cite{frequency_lower_bound2}.

We consider the problem of approximating frequency moments in insertion-only streams for $k\ge 3$. For any constant $c$ we show an $O(n^{1-2/k}\log(n)\log^{(c)}(n))$ upper bound on the space complexity of the problem. Here $\log^{(c)}(n)$ is the iterative $\log$ function.
To simplify the presentation, we make the following assumptions: $n$ and $m$
are polynomially far; approximation error $\epsilon$ and parameter $k$ are
constants.
We observe a natural bijection between streams and special matrices.
Our main technical contribution is a non-uniform sampling method on matrices. We call our method a \emph{pick-and-drop
sampling}; it samples a heavy element (i.e., element $i$ with frequency $\Omega(F_k)$) with probability $\Omega(1/n^{1-2/k})$ and gives approximation $\tilde{f_i} \ge (1-\epsilon)f_i$. In addition, the estimations never exceed the real values, that is $ \tilde{f_j} \le f_j$ for all $j$. As a result, we reduce the space complexity of finding a heavy element to $O(n^{1-2/k}\log(n))$ bits.
We apply our method of recursive sketches \cite{recursive} and resolve the problem with $O(n^{1-2/k}\log(n)\log^{(c)}(n))$ bits. We do not try to optimize the
space complexity as a function of $\epsilon$.

\subsection*{Overview of Main Ideas}

Pick-and-drop sampling has been inspired by a very natural behavior of children.
We observed the following pattern: a child picks a toy,
briefly plays with it, then drops the toy and picks a new
one. This pattern is repeated until the child picks the favorite toy and keeps it for a long time.
Indeed, children develop algorithms for selectivity \cite{zsgjgjhnbvnvfxgdjdgj1974}.

To illustrate the pick-and-drop method by example, assume that $m=r* t$ where $r=\lceil n^{1/k}\rceil$ and consider $r\times t$ matrix $M$ with entries $m_{i,j} = p_{k(i-1) + j}$. For $m\le n$ we aim to solve the following promise
problem with probability $2/3$:

\begin{itemize}
\item Case $1$: all frequencies are either zero or one.

\item Case $2$: $z$ appears in every row of $M$ exactly once (thus $f_z = r$).
All other frequencies are either zero or one.
\end{itemize}

Consider the following sampling method. Pick $r$ i.i.d. random numbers
$I_1,\dots, I_r,$ where $I_i$ is uniformly distributed on $\{1,2,\dots,t\}$.
For each $i = 1\dots {r-1}$ we check if there is a duplicate of $m_{i,
I_i}$ in the row $i+1$.
If the duplicate is found then we output ``Case $2$'' and stop; otherwise we repeat the test for $i+1$.
That is, the $i$-th sample is ``dropped,'' and the $(i+1)$-th sample is ``picked''. We repeat this experiment $T$ times independently and output ``Case $1$'' if no duplicate is found. Note that if the input represents Case $1$ then our method will always output ``Case $1$.''
Consider Case
$2$ and observe that if $m_{i, I_i} = z$ then our method will output ``Case $2$''.
Indeed, since $z$ appears in every row, the duplicate of $z$ will be found. The probability to miss $z$
entirely is
\begin{equation}\label{ksdfkjsdfkjsdfkjsd}
\left(1-{1\over t}\right)^{rT}.
\end{equation}
Recall that $m\le n, m=rt, r=\lceil n^{1/k}\rceil$. If $T = O(n^{1-2/k})$ with sufficiently large constant then the probability of error $(\ref{ksdfkjsdfkjsdfkjsd})$ is smaller than $1/3$.
We conclude that our promise problem can be resolved with $O(n^{1-2/k}\log(n))$ space.
Note how our solution depends on $r$. In general, the matrix should be carefully chosen.

Unfortunately the distribution of the frequent element in the stream can be arbitrary. Also our algorithm must recognize ``noisy'' frequencies that are large but negligible.
Clearly, the sampling must be more intricate but, luckily, not by much. In particular, the following method works. We introduce a local counter for each sample that counts the number of times $m_{i,I_i}$ appears in the suffix of the $i$-th row (this counting method is used in \cite{ams} for the entire stream). We maintain a global sample (and a global counter) as functions of the local samples and counters. Initially the global sample is the local sample of the first row.
Under certain conditions, the global sample can be ``dropped.'' If this is the case then the local sample of the current row is ``picked'' and becomes the new global sample.
The global sample is ``dropped'' when the local counter exceeds the global one. Also, the global sample is dropped if the global counter does not grow fast enough.
We use function $\lambda q$ where $\lambda$ is a parameter and $q$ is the number of rows that the global counter survived. If the global counter is smaller than $\lambda q$ then the global sample is ``dropped.''

In our analysis we concentrate on the case when $1$ is the heavy element, but it is possible to repeat our arguments for any $i$.
Our main technical contribution is Theorem \ref{fffsdsfbvdbf} that claims that $1$ will be outputted with probability $\Omega({f_1\over t})$  for sufficiently large $f_1$. Interestingly, Theorem \ref{fffsdsfbvdbf} holds for arbitrary distributions of frequencies. In Theorem \ref{fffsdsfbvdbfeererer} we show that there exist $r,t, \lambda$ such that a bound similar to $(\ref{ksdfkjsdfkjsdfkjsd})$ holds. We combine our new method with \cite{recursive} and obtain our main result in Theorem \ref{lefljsdjfjsdfljkdsfjklfd}.

\section{Pick-and-Drop Sampling}\label{tryjjjyjyjjyjtjttj}

Let $M$ be a matrix with $r$ rows and $t$ columns and with entries
$m_{i,j} \in [n]$. For $i\in [r], j\in [t], l\in [n]$ define:
\begin{equation}\label{ljdsffjsdfkjlsdfl}
d_{i,j} =|\{j': j\le j'\le t, m_{i,j'}=m_{i,j}\}|,
\end{equation}
\begin{equation}\label{lrerrrejdsffjsdfkjlsdfl}
f_{l,i} = |\{j \in [t]: m_{i,j}
= l\}|,
\end{equation}
\begin{equation}\label{lrerrrejwefwerwerdsffjsdfkjlsdfl}
f_l = |\{(i,j): m_{i,j} = l\}|,
\end{equation}
\begin{equation}\label{lrerrrejwefwerwerfddfdfdsffjsdfkjlsdfl}
F_k = \sum_{l=1}^n f_l^k, G_k = F_k - f_1^k.
\end{equation}

Note that there is a bijection between $r\times t$ matrices $M$ and streams $D$ of size $r\times t$ with elements $p_{it+j} = m_{i,j}$ where the definitions $(\ref{kljdfdsdfddflkjsdfljdsfljdsfklj}), (\ref{kljdflkjsdfljdsfljdsfklj})$ and $(\ref{lrerrrejwefwerwerdsffjsdfkjlsdfl}), (\ref{lrerrrejwefwerwerfddfdfdsffjsdfkjlsdfl})$ define equivalent frequency vectors for a matrix and the corresponding stream. W.l.o.g, we will consider streams of size $r\times t$ for some $r,t$ and will interchange the notions of a stream and its corresponding matrix.

\noindent
Let $\{I_j\}_{j=1}^r$ be i.i.d. random variables with  uniform distribution on $[t]$.
Define for $i = 1,\dots, r$:
\begin{equation}
s_i = m_{i,I_i}, c_i = d_{i,I_i}
\end{equation}
Let $\lambda$ be a parameter.
Define the following recurrent random variables:
\begin{equation}
S_1 = s_1, C_1 = c_1, q_1=1.
\end{equation}
Also (for $i = 2, \dots r$) if
\begin{equation}\label{lgffghjghkkjjkkjjsdjsdlkjsdjklf}
(C_{i-1} < \max\{\lambda q_{i-1}, c_i\})
\end{equation}
then define
\begin{equation}\label{ljsdjsdlkjsdjklf}
S_i = s_i,C_i = c_i,q_{i}=1;
\end{equation}
otherwise, define
\begin{equation}\label{ljsdjsdlkjsdjkrffdfdfddflf}
S_i = S_{i-1},C_i = C_{i-1} + f_{S_i, i},q_{i}= q_{i-1}+1
\end{equation}

\begin{theorem}\label{fffsdsfbvdbf}
Let $M$ be a $r\times t$ matrix.
There exist absolute constants $\alpha, \beta$ such that if
\begin{equation}\label{lksdklnldflkldflkmlfdlkngf}
\alpha(\lambda r + {{G}_3\over \lambda t} + {{G}_2\over t}) \le f_1 \le \beta t
\end{equation}
then
\begin{equation}\label{kdsdfkjldflldfndsffdfdfjksdsdkjdfkg}
P(S_r = 1) \ge {f_1\over 2t}.
\end{equation}
\end{theorem}

\begin{proof}
Denote
$
Q = \{(i,j): m_{i,j} = 1\}.
$
For $(i,j)\in Q$ define
\begin{equation}\label{kdsdfkjldfddsfdgtfghttlldfndsffdfdfjksdsdkjdfkg}
T_{i,j} = \overline{(A_{i,j}\cup B_{i,j} \cup H_{i,j})},
\end{equation}
where for $i>1$:
\begin{equation}\label{kdsdfndsffdfdffgfghfgdkjdfkg}
A_{i,j} = {\left(( C_{i-1} \ge d_{i,j}) \cap \left(S_{i-1} \neq
1\right)\right)},
\end{equation}
for $i<r$:
\begin{equation}\label{kdsdfndsfsdgsdgfsdffdfdffgfghfgdkjdfkg}
B_{i,j} = \left(\bigcup_{h=i+1}^r\left(d_{i,j} +
\sum_{u=i+1}^{h-1}f_{1,u} < c_{h}\right)\right), \ \ \ \ \
\end{equation}
\begin{equation}\label{kdsdfndsfcvvfghfgsdgsdgfsdffdfdffgfghfgdkjdfkg}
H_{i,j} = \left({\bigcup_{h=i+1}^r\left(d_{i,j} +
\sum_{u=i+1}^{h-1}f_{1,u} < (h-i)\lambda\right)}\right),
\end{equation}
and $A_{1,j} = B_{r,j} = H_{r,j} = \emptyset$.
We have
\begin{equation}\label{eljjnnslkkghy}
((s_i=1) \cap (S_{i-1} \neq 1) \cap  \overline{A_{i,I_i}}) \subseteq ((s_i=1) \cap (C_{i-1} < c_i)) \subseteq
$$
$$
((S_i = 1) \cap (q_i = 1)).
\end{equation}
Consider the case when $S_i = 1$ and $q_i = 1$ and
$$
d_{i, I_i} + \sum_{u=i+1}^{h-1}f_{1,u} \ge \max(\lambda (h-i), c_h)
$$
for all
$h>i$. In this case $S_h$ will be defined by $(\ref{ljsdjsdlkjsdjkrffdfdfddflf})$ and not by $(\ref{ljsdjsdlkjsdjklf})$; in particular, $S_h = S_i = 1$. Therefore,
\begin{equation}\label{eldfgdffdghdhggfhjjnnslkkghy}
((S_i=1) \cap (q_i = 1) \cap \overline{ B_{i,I_i}} \cap \overline{H_{i,I_i}}) \subseteq
(\bigcap_{h= i}^r(S_{h} = 1)).
\end{equation}

\noindent
Define $V_1 = ((s_1=1) \cap T_{1,I_1})$ and, for $i>1$,
$
V_i = ((s_i=1) \cap (S_{i-1} \neq 1) \cap T_{i,I_i}).
$
If follows from $(\ref{eljjnnslkkghy}),(\ref{eldfgdffdghdhggfhjjnnslkkghy})$ that, for any $i\in [r]$:
\begin{equation}\label{eldfgdghstryffdghdhggfhjjnnslkkghy}
V_i\subseteq (S_r=1),
\end{equation}
\begin{equation}\label{elddffdddgghfgdghstryffdghdhggfhjjnnslkkghy}
V_i \cap V_j =
\emptyset.
\end{equation}
Thus,
\begin{equation}\label{eldfgdddddssaadghstryffdghdhggfhjjnnslkkghy}
\sum_{i=1}^rP(V_i) = P\left(\cup_{i=1}^r V_i\right) \le P(S_r=1).
\end{equation}
For any $i>1$:
$$
P(V_i) \ge P((s_i=1) \cap T_{i,I_i}) - P(s_i=S_{i-1}=1).
$$
Also,
$$
\sum_{i=2}^r P(s_i=S_{i-1}=1) \le \sum_{i=2}^r P((s_i=1)\cap (\cup_{h\neq i}(s_{h}=1))) \le
$$
$$
(\sum_{i=1}^r P(s_i=1))^2 =
 \left({f_1\over
t}\right)^2.
$$
For any fixed $(i,j)\in Q$ events $I_i = j$ and $T_{i,j}$ are independent.
Indeed, $A_{i,j}$ is defined by $\{S_{i-1}, C_{i-1}\}$ that, in turn, is defined by $\{I_1,\dots, I_{i-1}\}$.
Similarly, $B_{i,j}$ is defined by $\{I_{i+1},\dots, I_r\}$.
Note that $H_{i,j}$ is a deterministic event.
By definition, $\{I_1,\dots, I_{i-1}, I_{i+1},\dots, I_r\}$ are independent of $I_i$; thus event $I_i = j$ and $T_{i,j} = \overline{(A_{i,j}\cup B_{i,j} \cup H_{i,j})}$ are independent.
Thus,
\begin{equation}\label{eldfgddhfhgghkkjmkjnkkhfhjjnnslkkghy}
\sum_{i=1}^rP((s_i=1) \cap T_{i,I_i}) = \sum_{(i,j)\in Q}P((I_i=j) \cap T_{i,j})  =
$$
$$
\sum_{(i,j)\in Q}P(I_i=j)P(T_{i,j})={1\over t}\sum_{(i,j)\in Q}P(T_{i,j}).
\end{equation}

Thus,
$$
P(S_r=1) \ge
{1\over t}\sum_{(i,j)\in Q}P(T_{i,j}) -
\left({f_1\over t}\right)^2.
$$
Lemma \ref{lm:sdfljsdfj} implies that $\sum_{(i,j)\in
Q}P(T_{i,j}) \ge 0.8f_1$. Thus if $\beta < 0.3$ then:
$$
P(S_r=1) \ge {f_1\over t}(0.8 - {f_1\over t}) \ge {f_1\over 2t}.
$$
Here we only use the second part of $(\ref{lksdklnldflkldflkmlfdlkngf})$. The first part is used in the proof of Lemma \ref{lm:sdfljsdfj}.
\end{proof}

\begin{lemma}\label{lm:sdfljsdfj}
There exist absolute constants $\alpha, \beta$ such that $(\ref{lksdklnldflkldflkmlfdlkngf})$ implies
$$\sum_{(i,j)\in Q}P(T_{i,j})>0.8f_1.$$
\end{lemma}
\noindent
It follows from Lemmas \ref{lm:jdlfddddnsjdsjlka}, \ref{lm:jdlfdtutyudddnsjdsjlka}, \ref{lm:jdlfdtutyugfgffgdddnsjdsjlka} and the union bound
that there exists at least $0.97f_1$ pairs $(i,j) \in Q$ such that
$P(A_{i,j} \cup B_{i,j} \cup H_{i,j}) \le 0.02$.
Recall that $T_{i,j} = \overline{(A_{i,j} \cup B_{i,j} \cup H_{i,j})}$; the lemma follows.

\subsection{Events of type $A$}\label{lkjldsjkljsdkldfslkfdkf}
For $(i,j) \in Q$ s.t. $i>1$
and for $l>1$ define:
$$
Y_{l,(i,j)} = \one_{A_{i,j}}\one_{(S_{i-1}=l)},
$$
$$
Y_{l,i} = \sum_{j\in [t], (i,j)\in Q} Y_{l,(i,j)},
$$

$$
Y_l = \sum_{i=2}^rY_{l,i},
$$
$$
Y = \sum_{l=2}^n Y_l,
$$
\begin{fact}\label{lm:dfghhghjdfdfdfgfhvba} $C_i
\le {f_{S_i}}.$ Also, if $q_i=1$ then $C_i
\le {f_{S_i, i}}.$
\end{fact}

\begin{proof}
Follows directly from $(\ref{ljsdjsdlkjsdjklf}),(\ref{ljsdjsdlkjsdjkrffdfdfddflf}).$
It is sufficient to prove that, for any $i$, there exists a set
$Q_i$ such that $C_i = |Q_i|$ and simultaneously $Q_i$ is a subset
of $\{(i',j) : m_{i',j} = S_i, i' \le i\}$. We prove the above claim
by induction on $i$. For $i=1$ the claim is true since we can define
$Q_1 = \{(1,j) : j \ge I_1\}$. For $i>2$ the description of the
algorithm implies the following. If $q_i = 1$ then we can put $Q_i =
\{(i,j) : j \ge I_i\}$.  If $q_i > 1$ then define $Q_i = Q_{i-1} \cup \{(i,j)
: m_{i,j} = S_i\}.$ Note that in this case $S_i = S_{i-1}.$
The second part follows from the description of the algorithms: if $p_i=1$ then $C_i = c_i, S_i = s_i$ and $c_i
= d_{i,I_i}(s_i) \le {f_{s_i, i}}.$
\end{proof}
\begin{fact}\label{lm:dfdfdfdfgfhvba}

\

\begin{enumerate}
\item $ Y_{l,i} \le f_l,$
\item If $q_{i-1} =1$ then $Y_{l, i} \le f_{l,i-1}.$
\end{enumerate}
\end{fact}
\begin{proof} Let
$(i,j)\in Q$ be such that $d_{i,j} > f_l$; then:
$$
Y_{l,(i,j)}  = \one_{(C_{i-1} \ge d_{i,j})}\one_{(S_{i-1} = l)} = \one_{(f_l\ge C_{i-1})}\one_{(C_{i-1} \ge d_{i,j})}\one_{ (S_{i-1} = l)}.
$$
We use Fact
\ref{lm:dfghhghjdfdfdfgfhvba} for the last equality. Thus, $Y_{l,(i,j)} = 0.$
Definition of $d_{i,j}$ implies $|\{j: (i,j)\in Q, d_{i,j} \le f_l\}| \le f_l$ for any fixed $i$ and $l$. Thus,
$$
Y_{l,i}=\sum_{j\in [t], (i,j)\in Q} Y_{l,(i,j)} \le f_l.
$$
Part $2$ following by repeating the above arguments and using the second statement of Fact \ref{lm:dfghhghjdfdfdfgfhvba}.
\end{proof}

\begin{definition}\label{def:efgedhdhdfhdf}
Let $1\le r_1\le r_2 \le r$ and $l\in [n]$. Call a pair $[r_1, r_2]$ an $l$-epoch if
$$
\forall i = r_1,\dots,r_2 : \ \ S_{i} = l,
$$
and
$$
q_{r_1} = q_{r_2+1} = 1,
$$
and
$$
\forall i = r_1+1,\dots, r_2  :\ \ q_{i} = q_{i-1}+1  .
$$
\end{definition}

\begin{lemma}\label{lm:sddsjdlfnsjgfgfhhfhgdsjlka}
Let $[r_1, r_2]$ be an $l$-epoch.  If $r_2>r_1$
then $$r_2-r_1  \le
 {1\over \lambda}\sum_{i=r_1}^{r_2-1} f_{l,i}.$$
\end{lemma}
\begin{proof}
First, observe that
$
q_{r_2-1} = r_2-r_1.
$
Second, $q_i > 1$ implies that $S_i$ is defined by $(\ref{ljsdjsdlkjsdjkrffdfdfddflf})$ and not by $(\ref{ljsdjsdlkjsdjklf})$ for all $r_1<i\le r_2$.
In particular, $C_{r_1}\le f_{l,r_1}$ and for $r_1<i\le r_2$ we have $C_{i}= C_{i-1}+f_{l,i}$.
Thus,
$$
C_{r_2-1} \le \sum_{i=r_1}^{r_2-1} f_{l,i}.
$$
Third, $C_{r_2-1} \ge \lambda q_{r_2-1}$ since $(\ref{lgffghjghkkjjkkjjsdjsdlkjsdjklf})$ must be false for $i=r_2$. Therefore,
$$
r_2-r_1 = q_{r_2-1} \le {1\over \lambda}C_{r_2-1} \le
 {1\over \lambda}\sum_{i=r_1}^{r_2-1} f_{l,i}.
$$
\end{proof}

\begin{lemma}\label{lm:sddsjdlfnsjdsjlka}
$Y_l \le {f^2_l\over \lambda} + f_l.$
\end{lemma}
\begin{proof}
Observe that the set $\{i:S_i=l\}$ is a collection of disjoint $l$-epochs.
Recall that $Y_l = \sum_{i=2}^r Y_{l,i}$ and $Y_{l,i}$ is non-zero only if $S_{i-1}$ is equal to $l$.
Thus we can rewrite $Y_l$ as:
$$
Y_l=\sum_{(r_1,r_2) \text{is an \ }\\ l\text{-epoch}}\left(\sum_{i=r_1+1}^{r_2+1}Y_{l,i}\right).
$$

\noindent
For any epoch such that $r_2>r_1$ we have by Lemmas \ref{lm:dfdfdfdfgfhvba} and \ref{lm:sddsjdlfnsjgfgfhhfhgdsjlka}:
$$
\sum_{i=r_1+1}^{r_2} Y_{l,i} \le (r_2-r_1)f_l \le {f_l \over
\lambda}\sum_{i=r_1}^{r_2-1} f_{l,i}.
$$

\noindent
Since all epochs are disjoint we have
$$
Y_l = \sum_{(r_1<r_2) \text{is an \ }\\ l\text{-epoch}}\left(\sum_{i=r_1+1}^{r_2+1}Y_{l,i}\right) + \sum_{(r_1=r_2) \text{is an \ }\\ l\text{-epoch}}Y_{l,r_2+1}=
$$
$$ \sum_{(r_1<r_2) \text{is an \ }\\ l\text{-epoch}}\left(\sum_{i=r_1+1}^{r_2}Y_{l,i}\right) + \sum_{(r_1,r_2)  \text{is an \ }\\ l\text{-epoch}}Y_{l,r_2+1}\le
$$
$$
{f_l \over
\lambda}\sum_{(r_1<r_2) \text{is an \ }\\ l\text{-epoch}}\left(\sum_{i=r_1}^{r_2-1} f_{l,i}\right) + \sum_{(r_1,r_2) \text{is an \ }\\ l\text{-epoch}}f_{l,r_2+1}\le
$$
$$
{f_l^2\over \lambda} + f_l.
$$
\end{proof}

\begin{lemma}\label{lm:jfgvdfbfbdlfnsjdsjlka}
$P(Y_l>0) \le {f_l \over t}.$
\end{lemma}
\begin{proof}

Since $I_i$ are independent and $0\le {f_{l,i}\over
t} \le 1$ we can apply Fact \ref{fct:dffdfd}:
$$
P(\cap_{i=1}^r(m_{i,I_i} \neq l)) = \prod_{i=1}^r(1-{f_{l,i}\over
t}) \ge (1-{f_{l}\over t}).
$$
Thus,
\begin{equation}\label{sddfgdgdf}
P(Y_l> 0) \le P(\cup_{i=1}^r( m_{i,I_i} = l)) \le {f_l \over t}.
\end{equation}
\end{proof}

\begin{lemma}\label{lm:jdlfddddnsjdsjlka}
There exists an absolute constant $\alpha$ such that $(\ref{lksdklnldflkldflkmlfdlkngf})$ implies that
$P(A_{i,j})
\le 0.01$ for at least $0.99f_1$ pairs $(i,j) \in Q$.
\end{lemma}
\begin{proof}
>From Lemmas \ref{lm:sddsjdlfnsjdsjlka}, \ref{lm:jfgvdfbfbdlfnsjdsjlka}:
$$
E(Y_l) \le {f_l\over t}({f^2_l\over \lambda}+f_l),
$$
$$
E(Y) = \sum_{l=2}^n E(Y_l) \le {{G}_3\over \lambda t} + {{G}_2\over t}.
$$
If follows that $\sum_{(i,j)\in Q} \one_{A_{i,j}} = Y$. Recall that by $(\ref{lksdklnldflkldflkmlfdlkngf})$:
$$
|Q|=f_1 \ge \alpha({{G}_3\over \lambda t} + {{G}_2\over t}) \ge \alpha E(\sum_{(i,j)\in Q} \one_{A_{i,j}}).
$$
Fact \ref{fct:sdlksdlklsd}
implies that there exists an absolute constant $\alpha$ such that the lemma is true.
\end{proof}

\noindent
The following fact is a well known. For completeness we present the proof.
\begin{fact}\label{fct:dffdfd}
Let $\alpha_1,\dots, \alpha_r$ be real numbers in $[0,1]$. Then
$$\prod_{i=1}^r(1-\alpha_i) \ge 1-(\sum_{i=1}^r\alpha_i).$$
\end{fact}
\begin{proof}
If $\sum_{i=1}^r \alpha_i\ge 1$ then
$$
\prod_{i=1}^r(1-\alpha_i) \ge 0 \ge 1-(\sum_{i=1}^r\alpha_i).
$$
Thus we can assume that $\sum_{i=1}^r \alpha_i < 1$.
We will prove the claim by induction on $r$. For $r =2$ we obtain $(1-\alpha_1)(1-\alpha_2)
= (1-\alpha_1-\alpha_2x + \alpha_1\alpha_2) \ge
(1-\alpha_1-\alpha_2)$. For $r>2$, we have, by induction,
$$\prod_{i=1}^r(1-\alpha_i) \ge
(1-(\sum_{i=1}^{r-1}\alpha_i))(1-\alpha_r) \ge
1-(\sum_{i=1}^{r}\alpha_i).$$
\end{proof}

\begin{fact}\label{fct:sdlksdlklsd}
Let $X_1,\dots, X_u$ be a sequence of indicator random variables.
Let $S = \{i: P(X_i = 1) \le \nu\}$. If $E(\sum_{i=1}^u X_i) \le
\mu u$ then $|S| \ge (1-{\mu\over \nu})u$.
\end{fact}
\begin{proof}
Indeed, $$\mu u \ge \sum_{i\notin S} P(X_i=1) \ge \nu(u-|S|).$$
\end{proof}

\subsection{Events of type $B$}\label{kfdkjskjdflfdlgffgdsdas}

For $(i,j)\in Q$ let $Z_{(i,j)}= \one_{B_{i,j}}.$ Let
$Z=\sum_{(i,j)\in Q} Z_{(i,j)}$. We use arguments that are similar to the ones from the previous section.
To stress the similarity we abuse the notation and denote by $Y_{l,h,(i,j)}$ the
indicator of the event that $h>i+1$, $s_h=l$ and
$$
\left(d_{i,j}+\sum_{u=i+1}^{h-1} f_{1,u}\right) < c_h.
$$
Define $Y_{l,h} = \sum_{(i,j)\in Q}Y_{l,h,(i,j)}$, $Y_l =
\sum_{h=1}^rY_{l,h}$.
\begin{fact}\label{fct:sdflkdfdffdjasdljasdgasdasdg}
$Y_l \le f_l.$
\end{fact}
\begin{proof}
Repeating the arguments from Fact \ref{lm:dfdfdfdfgfhvba} we have $c_h\one_{s_h=l}\le f_{l,h}$ and thus $ Y_{l,h} \le  f_{l,h}.$
\end{proof}

\begin{fact}\label{fct:sdflkjasdljasdgasdasdg}
$P(Y_l>0) \le {f_l\over t}.$
\end{fact}
\begin{proof}
The proof is identical to the proof of Lemma
\ref{lm:jfgvdfbfbdlfnsjdsjlka}.
\end{proof}

\begin{lemma}\label{lm:jdlfdtutyugfgffgdddnsjdsjlka}
There exist absolute constants $\alpha, \beta$ such that $(\ref{lksdklnldflkldflkmlfdlkngf})$ implies that
$P(B_{i,j})
\le 0.01$ for at least $0.99f_1$ pairs $(i,j) \in Q$.
\end{lemma}
\begin{proof}
Denote $Y=\sum_{l=1}^nY_l$. If follows  that $Z\le
Y$ and $E(Z)\le E(Y)$.
By Facts \ref{fct:sdflkjasdljasdgasdasdg} and
\ref{fct:sdflkdfdffdjasdljasdgasdasdg} if follows that $E(Y_l)\le
{f_l^2\over t}$. Thus by $(\ref{lksdklnldflkldflkmlfdlkngf})$:
$$
E(Z) \le E(Y) \le {F_2\over t} = {G_2\over t} + f_1{f_1\over t} \le (\alpha + \beta)f_1.
$$
We repeat the arguments from Lemma \ref{lm:jdlfddddnsjdsjlka}.
\end{proof}

\subsection{Events of type $H$}\label{kljdflkjdsfsfkldfknkdffknsdkksd}

\begin{definition}\label{def: dkjfjrglrlttrg}
Let $U = \{u_1,\dots,u_t\}$ and $W=\{w_1,\dots,w_t\}$ be two
sequences of non-negative integers. Let $(i,j)$ be a pair such that $1\le i\le
t$ and $1\le j\le u_i$. Denote $(i,j)$ as a \emph{loosing} pair (w.r.t. sequences $U,W$) if there exists $h, i\le h\le t$ such that:
$$
-j + \sum_{s=i}^h(u_s-w_s) < 0.
$$
Denote any pair that is not a loosing pair as a a \emph{winning} pair.
\end{definition}

In this section we consider the following pair $(U,W)$ of sequences. For $i=1,\dots, r$
let $u_i = f_{1,i}$ and $w_i = \lambda$.
\begin{fact}\label{fct:dfgfhfdgdf}
If $(i,j)$ is a winning pair w.r.t. $(U,W)$ then $H_{i,j'}$ does not occur where $j'$ is such that $m_{i,j'}=1$ and $d_{i,j'}=f_{1,i} - j + 1$.
\end{fact}
\begin{proof}
By Definition \ref{def: dkjfjrglrlttrg}, for every $i\le h\le r$:
\begin{equation}\label{eq:dfdgfsdggf}
-j + \sum_{l=i}^h u_l \ge \sum_{l=i}^h w_l.
\end{equation}
Since $\sum_{l=i}^h w_i = (h-i+1)\lambda$ and $d_{i,j'} = f_{1,i} - j+1$ we have for every $i\le h\le r$:
$$
d_{i,j'} + \sum_{l=i+1}^h d_{l,1} = f_{i,1} - j + 1 + \sum_{l=i+1}^h f_{l,1} =
$$
$$
-j +1 + \sum_{l=i}^h u_l \ge -j + \sum_{l=i}^h u_l \ge \sum_{l=i}^h w_l = (h-i+1)\lambda.
$$
Substitute $h$ by $h-1$ (for $h>i$):
$$
d_{i,j'} + \sum_{l=i+1}^{h-1} d_{l,1}  \ge (h-i)\lambda.
$$
Thus $H_{i,j'}$ does not occur, by $(\ref{kdsdfndsfcvvfghfgsdgsdgfsdffdfdffgfghfgdkjdfkg})$.
\end{proof}

\begin{lemma}\label{lm:jdlfdtutyudddnsjdsjlka}
There exists an absolute constant $\alpha$ such that $(\ref{lksdklnldflkldflkmlfdlkngf})$ implies that
$H_{i,j}$ does not occur for at least $0.99f_1$ pairs $(i,j) \in W$.
\end{lemma}
\begin{proof}
By Lemma  \ref{fct:erergerger} there exist at least
$$
\sum_{i=1}^r (u_i - w_i)
$$
winning pairs $(i,j)$ w.r.t. the $(U,W)$.
Also, $\sum_{i=1}^r u_i = \sum_{i=1}^r f_{1,i}=f_1$ and $\sum_{i=1}^r w_i = \lambda r$.
Thus there exist at least
$f_1-\lambda r$
winning pairs $(i,j)$ w.r.t. the $(U,W)$.
In the statement of Fact \ref{fct:dfgfhfdgdf} the mapping from $j$ to $j'$ is a bijection;
thus there exist at least $f_1-\lambda r$
pairs $(i,j')$ s.t. $m_{i,j'}=1$ and $H_{i,j'}$ does not occur.
By $(\ref{lksdklnldflkldflkmlfdlkngf})$ we have $f_1 \ge \alpha \lambda r$ and the lemma follows.
\end{proof}

\begin{definition}\label{def: dkjfjrbfghfghgglrlttrg}
Let $U = \{u_1,\dots,u_t\}$ and $W=\{w_1,\dots,w_t\}$ be two
sequences of non-negative integers. Let $1\le h < t.$ Let $U',W'$ be two sequences of size $t-h$
defined by $p'_i = u_{i+h}$, $q'_i = w_{i+h}$ for $i=1,\dots, t-h$.
Denote $U',W'$ as $h$-tail of the sequences $U,W$.
\end{definition}

\noindent
\begin{fact}\label{fct:rtheyjtyjukyukyuk}
If $(i,j)$ is a winning pair w.r.t. $h$-tail of $U,W$ then
$(i+h,j)$ is a winning pair w.r.t. $U,W$.
If $(i,j)$ is a winning pair w.r.t. $h$-tail of $U,W$ then
$(i,j)$ is a winning pair w.r.t. $U,W$.
\end{fact}
\begin{proof}
Follows directly from Definitions \ref{def: dkjfjrglrlttrg} and \ref{def: dkjfjrbfghfghgglrlttrg}.
\end{proof}

\begin{lemma}\label{fct:erergerger}
If $\sum_{s=1}^t(u_s-w_s) > 0$ then there exist at least $\sum_{s=1}^t(u_s-w_s)$ winning pairs.
\end{lemma}
\begin{proof}

We use induction on $t$. For $t=1$, any pair $(1,j)$
is winning if $1\le j\le u_1-w_1$. Consider $t>1$ and apply the following case analysis.

\begin{enumerate}
\item Assume  that there exist $1\le h< t$ such that $\sum_{s=1}^h(u_s-w_s) \le 0$. Consider the $h$-tail of $U,W$. By
induction and by Fact \ref{fct:rtheyjtyjukyukyuk}, there exist at least
$\sum_{s=h+1}^t(u_s-w_s) \ge \sum_{s=1}^t(u_s-w_s) $ winning pairs w.r.t. $U,W$.

\item Assume  that $(1,u_1)$ is a winning pair; it follows that $(1,j),\ j<u_1$ is a winning pair as well. If $\sum_{s=2}^t(u_s-w_s) > 0$ then, by
induction and by Fact \ref{fct:rtheyjtyjukyukyuk}, there exist at least
$\sum_{s=2}^t(u_s-w_s)$ winning pairs of the form $(i,j)$ where $i>1$. In total there are $u_1 + \sum_{s=2}^t(u_s-w_s) \ge \sum_{s=1}^t(u_s-w_s)$ winning pairs w.r.t. $U,W$.  The case when $\sum_{s=2}^t(u_s-w_s) < 0$ is trivial.

\item Assume that $(1), (2)$ do not hold. Then $u_1>0$. Indeed otherwise $u_1-w_1\le 0$ and thus $(1)$ is true. Also  $(1,1)$ is a winning pair. Indeed, otherwise there exists $1\le h < t$ such that $-1 + \sum_{i=1}^h(u_i-w_i)<0$. All numbers are integers thus $\sum_{i=1}^h(u_i-w_i)\le 0$ and $(1)$ is true.
Thus $(1,1)$ is a winning pair and $(1,u_1)$ is not a winning pair (by $(2)$). Therefore there exist $1<u\le u_1$ such that $(1,u-1)$ is a winning pair and $(1,u)$ is not a winning pair. In particular, there exists $1\le h<t$ such that
$$
-u + \sum_{s=1}^h(u_s-w_s) <0.
$$
On the other hand $(1,u-1)$ is a winning pair thus
$$
0\le 1-u + \sum_{s=1}^h(u_s-w_s).
$$
All numbers are integers and thus we conclude that
$$
\sum_{s=1}^h(u_s-w_s) = u-1.
$$
Consider the $h$-tail of $U,W$.
By induction, there exists at least
$$
\sum_{i=h+1}^{t}(u_i-w_i) = \sum_{i=1}^{t}(u_i-w_i) - (u-1)
$$
winning pairs w.r.t. the $h$-tail of $U,W$. By Fact \ref{fct:rtheyjtyjukyukyuk} there exist at least as many winning pairs w.r.t. $U,W$ of the form $(i,j)$ where $i>1$. By properties of $u$ there exist additional $(u-1)$ winning pairs of the form $(1,j), j\le u-1$.  Summing up we obtain the fact.
\end{enumerate}
\end{proof}

\section{The Streaming Algorithm}\label{sfdnnkjnksdndssdbffdgg}

\begin{fact}\label{fct:sdgsddegdfsg}
Let $v_1,\dots, v_n$ be a sequence of non-negative numbers and let
$k>2$. Then
$$
\left(\sum_{i=1}^n v_i^2\right)^{(k-1)} \le \left(\sum_{i=1}^n
v_i^k\right) \left(\sum_{i=1}^n v_i\right)^{(k-2)}
$$
\end{fact}
\begin{proof}
Define $\lambda_i = {v_i\over \sum_{j=1}^n v_j}$. Since
$g(x)=x^{k-1}$ is convex on the interval $[0,\infty)$ we can apply
Jensen's inequality and obtain:
$$
\left({\sum_{i=1}^n v_i^2\over \sum_{i=1}^n
v_i}\right)^{(k-1)}=(\sum_{i=1}^n \lambda_iv_i)^{(k-1)} \le
(\sum_{i=1}^n \lambda_iv_i^{(k-1)})={\sum_{i=1}^n v_i^k\over
\sum_{i=1}^n v_i}.
$$
\end{proof}

Let $D$ be a stream. Define
\begin{equation}\label{kbrgefgfftyutruturererredfskjcsfkjbksadfdjkbadsf}
\psi = {n^{1-(1/k)}G_k^{1/k}\over F_1} ,
\delta = 2^{\left\lceil 0.5\log_2(\psi)\right\rceil},
t =  \left\lceil {\delta F_1\over n^{1/k}} \right\rceil,
\lambda = \left\lceil {F_1\delta^3\over n} \right\rceil,
\end{equation}
where we use $(\ref{kljdfdsdfddflkjsdfljdsfljdsfklj})$ to define $F_k$.
We will make the following assumptions:
\begin{equation}\label{sdvsdfdsflkjkljklkjjlsdfsdff}
f_1 \le 0.1F_1, \ \ \ \ \ t\le F_1, \ \ \ \ F_1 (\mod t) = 0.
\end{equation}
Then it is possible to define a matrix
a $r\times t$ matrix $M$, where $r= F_1/t$ and  with entries $m_{i,j} = p_{ir+j}$.

\begin{fact}\label{fct:dffdhghggfhfdgdfffdff}
$1 \le \delta \le 2n^{(k-1)/2k.}$
\end{fact}
\begin{proof}
Indeed, $G_1 \le G_k^{1/k}n^{1-1/k}$
by H\"{o}lder
inequality and since $f_1 \le 0.1 F_1$ by $(\ref{sdvsdfdsflkjkljklkjjlsdfsdff})$ we have $\psi\ge 0.5$; thus, $\lceil 0.5\log_2(\psi)\rceil \ge 0$ and the lower bound follows. Also, $F_k^{1/k}$ is the $L_k$ norm for the frequency vector since since all frequencies are non-negative. Since $L_k \le L_1$
we conclude that $\psi \le n^{1-1/k}$ and the fact follows.
\end{proof}
Observe that there exists a frequency vector with $\delta = O(1)$: put $f_j = 1$ for all $i\in [n]$. At the same time there exists a vector with $\delta = \Omega(n^{(k-1)/2k})$:
put $f_1 = n$ and $f_j = 1$ for $j>2$. It is not hard to see that if $\delta$ is sufficiently large then a na\"{i}ve sampling method will find a heavy element.
For example, in the latter case, the heavy element occupies half of the stream.

\begin{fact}\label{fct:dfgfhfdgdfffdff}
$ \lambda r\le 4{G}_k^{1/k}$.
\end{fact}
\begin{proof}
Recall that $F_1 = rt$. The fact follows from the definitions of $\lambda$ and $t$.\end{proof}

\begin{fact}\label{fct:sdljsadgkjsdaf}
$$
{{G}_2\over t} \le {G}_k^{1/k}.
$$
\end{fact}
\begin{proof}
Define $\alpha = {k-3\over 2(k-2)}$. We have by H\"{o}lder
inequality:
\begin{equation}\label{kljdsakjkdkjldskjk}
{G}_2^{\alpha} \le {G}_k^{2\alpha\over k}n^{\alpha(1-{2\over k})} = {G}_k^{k-3\over k(k-2)}n^{{k-3\over 2k}}.
\end{equation}
Also, by Fact \ref{fct:sdgsddegdfsg}
\begin{equation}\label{kljfdhhfgfgdsakjkdkjldskjk}
{G}_2^{1-\alpha} = {G}_2^{k-1\over 2(k-2)} \le {G}_k^{1\over 2(k-2)}
{G}_1^{1\over 2}.
\end{equation}
Thus,
$$
{G}_2\le {G}_k^{k-3\over k(k-2)}n^{{k-3\over 2k}}{G}_k^{1\over 2(k-2)}
F_1^{1\over 2} =
$$
$$
{G}_k^{1\over k}{F_1\over n^{1/k}}\left({{G}_k^{1\over k}n^{{k-1\over
k}}\over F_1}\right)^{1/2} = t{G}_k^{1\over k}.
$$
\end{proof}

\begin{fact}\label{fct:sdfasdasdsadfgsda}
${{G}_3 \over \lambda t}\le {G}_k^{1/k}$.
\end{fact}
\begin{proof}
By H\"{o}lder inequality,
\begin{equation}\label{kljdfgdffyhfdsakjkdkjldsnljklkjk}
{G}_3 \le {G}_k^{3/k}n^{1-(3/k)}.
\end{equation}
Thus
$$
{{G}_3\over \lambda t} =  {n^{1+(1/k)}{G}_3 \over F_1^2\delta^4} \le {n^{2-(2/k)}{G}_k^{3/k} \over
F_1^2\delta^4} \le {G}_k^{1/k}.
$$
\end{proof}

\begin{theorem}\label{fffsdsfbvdbfeererer}
Let $M$ be a $r\times t$ matrix such that $(\ref{kbrgefgfftyutruturererredfskjcsfkjbksadfdjkbadsf})$
is true.
Then there exist absolute constants $\alpha, \beta$ such that
\begin{equation}\label{kdsdfkjndsddddffffdfdfjksdsdkjdfkg}
\alpha G_k^{1/k} \le f_1 \le \beta t
\end{equation}
imply
\begin{equation}\label{kdsdcffkjbnbmmnmnldflldfndsffdfdfjksdsdkjdfkg}
P(S_r = 1) \ge {\delta\over 2n^{1-(2/k)}}.
\end{equation}
\end{theorem}
\begin{proof}
By $(\ref{kdsdfkjndsddddffffdfdfjksdsdkjdfkg})$ and Facts \ref{fct:sdfasdasdsadfgsda}, \ref{fct:sdljsadgkjsdaf}, \ref{fct:dfgfhfdgdfffdff}:
$$
6\alpha (\lambda r + {{G}_3\over \lambda t} + {{G}_2\over t}) \le f_1 \le \beta t.
$$
Also, $(\ref{kbrgefgfftyutruturererredfskjcsfkjbksadfdjkbadsf})$ implies $f_1/t \ge {\delta\over n^{1-(2/k)}}$. Thus, $(\ref{kdsdcffkjbnbmmnmnldflldfndsffdfdfjksdsdkjdfkg})$ follows from Theorem \ref{fffsdsfbvdbf}.
\end{proof}

\noindent Algorithm \ref{approxzufgjgfyyjgyjztfgfggfikfinder} describes our implementation of the pick-and-drop sampling.
\begin{algorithm}
\caption{P\&D$(M,r,t, \lambda)$}
\label{approxzufgjgfyyjgyjztfgfggfikfinder}
\begin{algorithmic}

\State Generate i.i.d. r.v.  $\{I_j\}_{j=1}^r$
with  uniform distribution on $[t]$.

\State $S_1 = m_{1,I_1},$
\State $C_1 = d_{1,I_1},$
\State $q_1=1$.

\For {$i=2\to r$}

\State compute $s_i = m_{i,I_i}, c_i = d_{i,I_i}$

\If   {$(C_{i-1} < \max\{\lambda q_{i-1}, c_i\})$}
\State $S_i = s_i,$
\State $C_i = c_i,$
\State $q_{i}=1$

\Else
\State $S_i = S_{i-1},$
\State  $C_i = C_i + f_{S_i, l},$
\State  $q_{i}= q_{i-1}+1$
\EndIf
\EndFor

\State {\bf Output $(S_r, C_r)$.}

\end{algorithmic}
\end{algorithm}

\begin{theorem}\label{dsksdkhfkhdsfkhdskhfkdfs}
Denote $f_i^k > 100\sum_{j\neq i} f_j^k$ as a \emph{heavy} element.
There exist a (constructive) algorithm that makes one pass over the stream
and uses $O(n^{1-2/k}\log(n))$ bits. The algorithm outputs a pair $(i, \tilde{f}_i)$ such that $\tilde{f}_i \le f_i$ with probability $1$. If there exists a heavy element $f_i$ then also with constant probability the algorithm will output $(i,\tilde{f}_i)$ such that $(1- \epsilon)f_i \le \tilde{f}_i$.
\end{theorem}
\begin{proof}
Define $t$ as in $(\ref{kbrgefgfftyutruturererredfskjcsfkjbksadfdjkbadsf})$. W.l.o.g., we can assume that $F_1$ is divisible by $t$.
Note that if $t> F_1$ or $f_1 \ge 0.1F_1$ then it is possible to find a heavy element with $O(n^{1-2/k})$ bits by existing methods such as \cite{684566}.
Otherwise, a stream $D$ defines a matrix $M$ for which we compute $O(n^{1-2/k}/ \epsilon\delta)$ independent pick-and-drop samples.
Since we do not know the value of $\delta$ we should repeat the experiment for all possible values of $\delta$.
Output the element with the maximum frequency.
With constant probability the output of the pick-and-drop sampling will include a
$(1-\epsilon)$ approximation of the frequency $f_i$. Thus, there will be no other $f_j$ that can give a larger approximation and replace a heavy element.
The total space will define geometric series that sums to $O(n^{1-2/k}\log(n))$.

If we know $F_1$ ahead of time then we can compute the value of $t$ for any possible $\delta$ and thus solve the problem in one pass. However, one can show that the well-known doubling technique (when we double our parameter $t$ each time the size of the stream doubles) will work in our case and thus one pass is sufficient even without knowing $F_1.$
\end{proof}

\noindent
Recall that in \cite{recursive} we developed a method of recursive sketches with the following property:
given an algorithm that finds a heavy element and uses memory $\mu(n)$, it is possible to solve the frequency moment problem in space $O(\mu(n)\log^{(c)}(n))$.
In \cite{recursive} we applied recursive sketches with the method of Charikar et.al. \cite{684566}.
Thus, we can replace the method from \cite{684566} with Theorem \ref{dsksdkhfkhdsfkhdskhfkdfs} and obtain:
\begin{theorem}\label{lefljsdjfjsdfljkdsfjklfd}
Let $\epsilon$ and $k$ be constants.
There exists a (constructive) algorithm that computes $(1\pm \epsilon)$-approximation of $F_k$, uses $O(n^{1-2/k}\log(n)\log^{(c)}(n))$ memory bits, makes one pass and errs with probability at most $1/3$.
\end{theorem}

\bibliographystyle{plain}
\bibliography{Bibliography}

\begin{thebibliography}{10}

\bibitem{ams}
Noga Alon, Yossi Matias, and Mario Szegedy.
\newblock The space complexity of approximating the frequency moments.
\newblock {\em J. Comput. Syst. Sci.}, 58(1):137--147, 1999.

\bibitem{DBLP:conf/focs/AndoniKO11}
Alexandr Andoni, Robert Krauthgamer, and Krzysztof Onak.
\newblock Streaming algorithms via precision sampling.
\newblock In {\em FOCS}, pages 363--372, 2011.

\bibitem{frequency_lower_bound1}
Ziv Bar-Yossef, T.~S. Jayram, Ravi Kumar, and D.~Sivakumar.
\newblock An information statistics approach to data stream and communication
  complexity.
\newblock {\em J. Comput. Syst. Sci.}, 68(4):702--732, 2004.

\bibitem{711822}
Ziv Bar-Yossef, T.~S. Jayram, Ravi Kumar, D.~Sivakumar, and Luca Trevisan.
\newblock Counting distinct elements in a data stream.
\newblock In {\em RANDOM '02: Proceedings of the 6th International Workshop on
  Randomization and Approximation Techniques}, pages 1--10, London, UK, 2002.
  Springer-Verlag.

\bibitem{1250891}
Paul Beame, T.~S. Jayram, and Atri Rudra.
\newblock Lower bounds for randomized read/write stream algorithms.
\newblock In {\em STOC '07: Proceedings of the thirty-ninth annual ACM
  symposium on Theory of computing}, pages 689--698, New York, NY, USA, 2007.
  ACM.

\bibitem{recursive}
Vladimir Braverman and Rafail Ostrovsky.
\newblock Recursive sketching for frequency moments.
\newblock {\em CoRR}, abs/1011.2571, 2010.

\bibitem{1374470}
Amit Chakrabarti, Graham Cormode, and Andrew McGregor.
\newblock Robust lower bounds for communication and stream computation.
\newblock In {\em STOC '08: Proceedings of the 40th annual ACM symposium on
  Theory of computing}, pages 641--650, New York, NY, USA, 2008. ACM.

\bibitem{frequency_lower_bound2}
Amit Chakrabarti, Subhash Khot, and Xiaodong Sun.
\newblock Near-optimal lower bounds on the multi-party communication complexity
  of set disjointness.
\newblock In {\em IEEE Conference on Computational Complexity}, pages 107--117,
  2003.

\bibitem{684566}
Moses Charikar, Kevin Chen, and Martin Farach-Colton.
\newblock Finding frequent items in data streams.
\newblock In {\em ICALP '02: Proceedings of the 29th International Colloquium
  on Automata, Languages and Programming}, pages 693--703, London, UK, 2002.
  Springer-Verlag.

\bibitem{frequency_impr2}
Don Coppersmith and Ravi Kumar.
\newblock An improved data stream algorithm for frequency moments.
\newblock In {\em SODA}, pages 151--156, 2004.

\bibitem{776778}
Graham Cormode, Mayur Datar, Piotr Indyk, and S.~Muthukrishnan.
\newblock Comparing data streams using hamming norms (how to zero in).
\newblock {\em IEEE Trans. on Knowl. and Data Eng.}, 15(3):529--540, 2003.

\bibitem{796530}
Joan Feigenbaum, Sampath Kannan, Martin Strauss, and Mahesh Viswanathan.
\newblock An approximate l1-difference algorithm for massive data streams.
\newblock In {\em FOCS '99: Proceedings of the 40th Annual Symposium on
  Foundations of Computer Science}, page 501, Washington, DC, USA, 1999. IEEE
  Computer Society.

\bibitem{5215}
Philippe Flajolet and G.~Nigel Martin.
\newblock Probabilistic counting algorithms for data base applications.
\newblock {\em J. Comput. Syst. Sci.}, 31(2):182--209, 1985.

\bibitem{frequency_impr1}
Sumit Ganguly.
\newblock Estimating frequency moments of data streams using random linear
  combinations.
\newblock In {\em APPROX-RANDOM}, pages 369--380, 2004.

\bibitem{DBLP:journals/corr/abs-1104-4552}
Sumit Ganguly.
\newblock Polynomial estimators for high frequency moments.
\newblock {\em CoRR}, abs/1104.4552, 2011.

\bibitem{DBLP:journals/corr/abs-1201-0253}
Sumit Ganguly.
\newblock A lower bound for estimating high moments of a data stream.
\newblock {\em CoRR}, abs/1201.0253, 2012.

\bibitem{1459774}
Sumit Ganguly and Graham Cormode.
\newblock On estimating frequency moments of data streams.
\newblock In {\em APPROX '07/RANDOM '07: Proceedings of the 10th International
  Workshop on Approximation and the 11th International Workshop on
  Randomization, and Combinatorial Optimization. Algorithms and Techniques},
  pages 479--493, Berlin, Heidelberg, 2007. Springer-Verlag.

\bibitem{stable}
Piotr Indyk.
\newblock Stable distributions, pseudorandom generators, embeddings, and data
  stream computation.
\newblock {\em J. ACM}, 53(3):307--323, 2006.

\bibitem{frequency}
Piotr Indyk and David Woodruff.
\newblock Optimal approximations of the frequency moments of data streams.
\newblock In {\em STOC '05: Proceedings of the thirty-seventh annual ACM
  symposium on Theory of computing}, pages 202--208, New York, NY, USA, 2005.
  ACM.

\bibitem{1265565}
T.~S. Jayram, Andrew McGregor, S.~Muthukrishnan, and Erik Vee.
\newblock Estimating statistical aggregates on probabilistic data streams.
\newblock In {\em PODS '07: Proceedings of the twenty-sixth ACM
  SIGMOD-SIGACT-SIGART symposium on Principles of database systems}, pages
  243--252, New York, NY, USA, 2007. ACM.

\bibitem{Jayram:2011:OBJ:2133036.2133037}
T.~S. Jayram and David Woodruff.
\newblock Optimal bounds for johnson-lindenstrauss transforms and streaming
  problems with sub-constant error.
\newblock In {\em Proceedings of the Twenty-Second Annual ACM-SIAM Symposium on
  Discrete Algorithms}, SODA '11, pages 1--10. SIAM, 2011.

\bibitem{nelson}
Daniel~M. Kane, Jelani Nelson, and David~P. Woodruff.
\newblock On the exact space complexity of sketching and streaming small norms.
\newblock In {\em Proceedings of the 21st Annual ACM-SIAM Symposium on Discrete
  Algorithms (SODA 2010)}, 2010.

\bibitem{1807094}
Daniel~M. Kane, Jelani Nelson, and David~P. Woodruff.
\newblock An optimal algorithm for the distinct elements problem.
\newblock In {\em PODS '10: Proceedings of the twenty-ninth ACM
  SIGMOD-SIGACT-SIGART symposium on Principles of database systems of data},
  pages 41--52, New York, NY, USA, 2010. ACM.

\bibitem{1496816}
Ping Li.
\newblock Compressed counting.
\newblock In {\em SODA '09: Proceedings of the Nineteenth Annual ACM -SIAM
  Symposium on Discrete Algorithms}, pages 412--421, Philadelphia, PA, USA,
  2009. Society for Industrial and Applied Mathematics.

\bibitem{strbook}
S.~Muthukrishnan.
\newblock Data streams: algorithms and applications.
\newblock {\em Found. Trends Theor. Comput. Sci.}, 1(2):117--236, 2005.

\bibitem{1807101}
Jelani Nelson and David~P. Woodruff.
\newblock Fast manhattan sketches in data streams.
\newblock In {\em PODS '10: Proceedings of the twenty-ninth ACM
  SIGMOD-SIGACT-SIGART symposium on Principles of database systems of data},
  pages 99--110, New York, NY, USA, 2010. ACM.

\bibitem{zsgjgjhnbvnvfxgdjdgj1974}
Anne~D. Pick and Gusti~W. Frankel.
\newblock A developmental study of strategies of visual selectivity.
\newblock {\em Child Development}, 45(4):pp. 1162--1165, 1974.

\bibitem{982817}
David Woodruff.
\newblock Optimal space lower bounds for all frequency moments.
\newblock In {\em SODA '04: Proceedings of the fifteenth annual ACM-SIAM
  symposium on Discrete algorithms}, pages 167--175, 2004.

\bibitem{DBLP:reference/db/Woodruff09}
David~P. Woodruff.
\newblock Frequency moments.
\newblock In {\em Encyclopedia of Database Systems}, pages 1169--1170. 2009.

\bibitem{Woodruff:2012:TBD:2213977.2214063}
David~P. Woodruff and Qin Zhang.
\newblock Tight bounds for distributed functional monitoring.
\newblock In {\em Proceedings of the 44th symposium on Theory of Computing},
  STOC '12, pages 941--960, New York, NY, USA, 2012. ACM.

\end{thebibliography}

\end{document}